\documentclass[a4paper]{article}
\usepackage{amsmath,amssymb,amsthm,amssymb,graphicx}

\begin{document}

\title{An analog of Heisenberg uncertainty relation in 
prequantum classical field theory}

\author{Andrei Khrennikov\\ International Center for Mathematical Modelling
\\in Physics and Cognitive Sciences,\\
University of V\"axj\"o,  V\"axj\"o, Sweden\\
Institute of Information Security\\
Russian State University for Humanities, Moscow, Russia\\
Andrei.Khrennikov@vxu.se}

\maketitle

\begin{abstract}
Prequantum classical statistical field theory (PCSFT) is a model which provides a possibility 
to represent averages of quantum observables, including correlations of observables on subsystems of a composite system,
 as averages with respect to fluctuations of classical random fields. 
PCSFT is a classical model of the wave type. For example, ``electron'' is 
described by electronic field. In contrast to QM, this field is a real physical field and not a field of probabilities.
An important point is that the prequantum field of e.g. electron
contains the irreducible contribution of the background field, vacuum fluctuations.
In principle, the traditional QM-formalism can be considered as a special regularization procedure: subtraction of averages 
with respect to vacuum fluctuations. In this paper
we derive a classical analog of the Heisenberg-Robertson inequality for dispersions of functionals of classical (prequantum)
fields. PCSFT   Robertson-like inequality provides a restriction on the product of classical dispersions. However, this restriction 
is not so rigid as in QM. The quantum dispersion corresponds to the difference between e.g. the electron field dispersion and  the
dispersion of vacuum fluctuations. Classical Robertson-like inequality contains these differences. Hence, it does not imply  
such a rigid estimate from below for dispersions as it was done in QM.
\end{abstract}

\section{Introduction}

At the very beginning of QM 
the idea that quantum mechanics is simply a special model of {\it wave mechanics} was quite popular. It was supported by 
the discovery of ``quantum wave equation'' by Schr\"odinger and by the association with each particle its wave-length, the De Broglie 
wave-length. However, in spite of increasable success in ``technical applications''\footnote{Concrete spectra were found by using 
Schr\"odinger's representation.}, quantum wave mechanics 
was  ideologically inconsistent. Both Schr\"odinger and De Broglie should give up in front of the difficulties of the interpretation of 
``quantum waves'' as real physical waves. Although both Schr\"odinger and De Broglie did not like the Copenhagen interpretation, 
they were not able to present a consistent ``physical waves approach'' to QM.\footnote{In particular, De Broglie's double solution model was not so attractive even for its creator. He was happy with Bohmian mechanics, but the latter has its own difficulties, e.g., nonlocality.} Finally, the interpretation of quantum waves as waves of probability (proposed by Born) became commonly accepted. 

One of the main problems of the ``physical wave interpretation'' was the impossibility to 
describe a composite system by waves defined on the physical space, $X = {\bf R}^3.$ 
The wave function of a composite system is 
defined on the space $X_m = {\bf R}^{3m},$ where $m$ is the number of subsystems. Pauli wrote that one may consider physical waves,
but they will be defined on unphysical space. In particular, this problem was the main reason why Schr\"odinger 
accepted the probabilistic interpretation of the wave function, see \cite{AKK}--\cite{AKK6} for recent debates.

Nevertheless, after 80 years of the dominance of the Copenhagen interpretation, nowadays various wave models are very popular in 
attempts to go beyond QM. At the present time the most successful models are stochastic electrodynamics (SED), see, 
e.g., \cite{R2}--\cite{Teo}, and semiclassical theory, e.g., \cite{C1}--\cite{C3}. 
Some ideas of SED are quite similar to those which will be discussed in this paper. A crucial point of SED 
is the assumption on the presence of the so called background field (zero point field, field of quantum fluctuations). This idea is very 
physical. Everybody agrees that the ``totally empty space'' is only a mathematical idealization. By SED there are no 
``free quantum particles''. It is impossible to isolate, e.g., an electron from the zero point field. Dynamics of quantum systems is a 
motion in the sea of quantum fluctuations. The latter produce new dynamical and statistical effects which are known as quantum effects. 
By SED these effects are purely classical. The mystery of these effects is in the use of incomplete information, 
cf. Einstein \cite{E}--\cite{EPR}, ignoring of the zero point field, cf. SED.

Before going to my own wave model, I would like to present motivations to go beyond QM, 
to create prequantum models emerging QM. One of the standard questions after my talks is ``What is the reason 
to put efforts into such an activity? QM works very well!''

The main reason, see Einstein \cite{E}--\cite{EPR}, is creation a finer description of micro-processes than given by the wave function.  
Such a description is based on new parameters (``hidden variables'')  providing a possibility for monitoring of an individual quantum system. 
We call this project ``the great Einstein's dream''. At the very beginning of the quantum epoch Einstein dreamed of a kind of Hamiltonian 
dynamics for quantum particles. It seems that he never gave up and dreamed of a deterministic prequantum model until the last days of his 
life. However, in the 1930s he concentrated efforts to show that quantum randomness is reducible to classical randomness, i.e., 
quantum statistics can be reproduced in the classical probabilistic framework, see his correspondence with Schr\"odinger \cite{SCH}.

We know that even in the classical world deterministic dynamics is not so common. Theory of stochastic processes 
(including stochastic differential equations) is widely used in classical physics and other domains of science.
Personally I believe that the great Einstein's dream was wrong. However, Einstein's dream of reduced quantum randomness, 
i.e., a possibility to describe behavior of quantum systems by classical stochastics seems to be true.

Although it is too early to predict experimental consequences of the realization of 
``the reduced Einstein's dream", ideological consequences are evident. We would create a harmonic and unified picture of 
physical reality: QM and classical statistical mechanics would be described by the same probabilistic model.     

My model PCSFT \cite{EKHR}--\cite{IZV}, is a realization of ``the reduced Einstein's dream'', cf. SED and semiclassical model, 
cf. Nelson's stochastic mechanics \cite{N} and its generalization by Davidson \cite{D1}, \cite{D2}, cf. also  
tomographic approach of Man'ko et al. \cite{M1}--\cite{M2}  reproducing all quantum statistical predictions by operating 
with classical probability.

To simplify mathematical presentation, we assume that all operators under consideration are bounded and moreover 
that they are of the trace class (so called nuclear operators by the terminology of functional analysis).

\section{Correspondence between terminologies of QM and PCSFT}

In \cite{EKHR}--\cite{IZV} it was proposed to describe an ensemble of ``quantum particles'' prepared in a state (maybe mixed) given (in QM-formalism) by 
the density operator $\rho$ by a random classical field which covariance operator coincides with $\rho$ (and mean value equals to zero). 
It seems that Gaussian fields provide the best matching with QM. However, it became evident only in the process of generalization of 
PCSFT to composite systems \cite{CS}. For a quantum observable given by a symmetric linear operator $\hat A,$ 
we introduce the corresponding classical variable
$$
f_A (\phi) = (A \phi, \phi),
$$
quadratic functional of classical fields. In the real physical model
 the argument $\varphi$ varies in a complex Hilbert space $H=L_2({\bf R}^3)$ of square-summable (complex valued) 
functions. In the mathematical formalism we proceed with an arbitrary complex Hilbert space $H.$ To escape mathematical 
difficulties related to theory of Gaussian measures on infinite-dimensional spaces, the reader can restrict considerations
to the case of finite-dimensional spaces, i.e., $H={\bf C}^n,$ where ${\bf C}$ is the field of complex numbers. 

The basic mathematical formula \cite{IZV} coupling QM-average and PCSFT-average is 
\begin{equation}
\label{BBPP}
\int_H f_A (\phi) d\mu_\rho (\phi) = {\rm Tr} \rho \hat A \equiv \langle \hat A \rangle_\rho,
\end{equation}
where $\mu_\rho$ is a probability measure with the covariance operator $\rho.$ 
By using the language of probability theory we can write this equality as
\begin{equation}
\label{a1}
E_{\mu_\rho} f_A = \langle \hat A \rangle_\rho,
\end{equation}
where $E_{\mu_\rho}$ is the classical probabilistic expectation. Although in our previous papers \cite{EKHR}--\cite{IZV}
 this formula was proved only 
for  symmetric operators, it is easy to check that it is also valid for an arbitrary (bounded) linear operator. We will use this mathematical 
fact at the very end of the paper. 

The next natural question is on coupling of the PCFT- and QM-dispersions 
and, hence,  on an analog of Heisenberg's uncertainty relation in PCSFT. 
``Old PCSFT'' \cite{EKHR}--\cite{IZV} did not provide a reasonable coupling 
between the classical and quantum dispersions. Heisenberg-like inequalities were not found, 
a role of noncommutativity was unclear. 

We clarify this problem. Let $\psi \in H, ||\psi|| = 1,$ be a pure quantum state. In QM it is represented by the density operator 
$$
\rho_\psi = \psi \psi\dagger.
$$ Consider a Gaussian random variable $\phi (\omega)$ valued in $H$ and having the covariance operator
 $\rho_\psi.$ It can be represented as $\phi (\omega) = \xi (\omega) \psi,$ where $\xi$ takes its values in the field of complex numbers 
 ${\bf C},$ it is scalar random variable, and it has zero mean value and dispersion 1. Then
$$
E (u, \phi (\omega)) (\phi (\omega), v) =  \langle \rho_\psi u \mid v \rangle.
$$
Let operator  $\hat A$ be self-adjoint and bounded. To simplify considerations, assume that its QM-average 
$\langle \hat A \rangle_\psi =0.$ By (\ref{a1}) its PCSFT-average 
$E_{\mu_{\rho_\psi}} f_A = 0.$ Thus $\sigma^2_{\mu_{\rho_\psi}} (f_A) = E_{\mu_{\rho_\psi}} f_A^2 = 
E \xi^4 (\omega) \langle \hat A \psi \mid \psi \rangle^2 = 3 \langle \hat A \rangle^2_\psi.$

Thus the PCSFT-dispersion has no coupling with the QM-dispersion $\sigma_\psi^2 (\hat A) = \langle \hat A^2 \rangle_\psi$ 
(we remind that it was assumed that $\langle \hat A \rangle_\psi = 0).$

Hence, the ``old PCSFT'' \cite{EKHR}--\cite{IZV}
 provides matching of averages, but not dispersions. It was a problem. Recently, PCSFT was succesfully 
generalized to composite quantum system, see \cite{CS}. Surprisingly, one can proceed without the 
tensor product state space. ``Quantum waves'' 
for composite systems can be described by the Cartesian product of Hilbert spaces (similar to the 
classical description of a few particles). One of the main reasons (at least for Schr\"odinger and Pauli) 
to support probabilistic interpretation of quantum waves 
 disappeared.  We now point to the most important lesson of
 the construction  which was used in \cite{CS} to extend PCSFT to composite systems.

\medskip

To construct a proper probability measure for a composite system, see \cite{CS}, one should change the correspondence between
density operators of QM and covariance operators of PCSFT even for a single system: not simply 
\begin{equation}
\label{b1}
\rho \to \rho, 
\end{equation}
but 
\begin{equation}
\label{b2}
\rho \to D_\rho = \rho + I.
\end{equation}
Thus the  quantum density operator $\rho$ is perturbed by the unit operator $I$. 
The latter is the covariance operator of white noise.

For a pure state $\psi,$ we set 
$$
D_{\rho_\psi}\equiv D_{\psi}.
$$

From this viewpoint,
 QM is a special mathematical formalism designed to eliminate effects of  vacuum fluctuations of the white noise type. 
It is a natural formalism to describe observations performed on the random background. The contribution of this background should 
be subtracted. We have for trace class operator  $\hat A$:
$$
E_{\mu_{D_\rho}} f_A = {\rm Tr} \rho \hat A + {\rm Tr \hat A},
$$ 
i.e.,
\begin{equation}
\label{a2}
\langle \hat A \rangle_\rho = E_{\mu_{D_\rho}} f_A - {\rm Tr} \hat A
\end{equation}
or at least formally (there are some mathematical difficulties in the case of the infinite-dimensional Hilbert space):
$$\langle \hat A \rangle_\rho = E_{\mu_{D_\rho}} f_A - E_{\mu_I} f_A.$$
QM-formalism can  be interpreted as a rather special {\it regularisation procedure.} 
If ${\rm Tr} \hat A = \infty,$ then PCSFT - average (with respect to $\mu_D$) is not defined: 
the Gaussian integral diverges. Of course, this effect is a consequence of the infinite-dimension of the
state space. However, QM-formalism provides its regularization.

Coupling (\ref{a2}) between QM- and PCSFT-averages is not so straight-forward as (\ref{a1}). 
However, as we will see, equality (\ref{a2}) will provide coupling between the QM- and PCSFT-dispersions.

We remark that correspondence (\ref{b2}) appeared originally by pure mathematical reasons. 
To construct a positively defined operator, on the Cartesian product of state spaces of subsystems of a composite system \cite{CS}
one should modify (\ref{b1}) to (\ref{b2}) even for each  subsystem. However, this modification has a natural physical 
interpretation. The background field of the white noise type should be taken into account. Its contribution was missed
in ``old PCSFT''  \cite{EKHR}--\cite{IZV}. New PCSFT taking into account so to say vacuum fluctuations became even closer to SED.
Although in reality this white noise exists, its contribution can be eliminated from all experimental averages, since both 
``quantum systems'' and measurement devices are located in the ``vacuum thermostat''. The formalism of QM eliminates 
from all answers average with respect to fluctuations of this thermostat. It is a good place to come back to the question
on possible consequences of creation of PCSFT. In particular, it will start to play a role when experimental technology 
will approach such a degree of precision that individual fluctuations of vacuum would be visible. At that level it would not be more possible 
just to subtract averages with respect to this fluctuations from all answers. The boundary of possible application of the formalism of QM 
will be approached.     

\section{ Coupling of dispersions.}

We restrict considerations to pure states.
Let $\hat A$ be symmetric. Its dispersion in a pure state $\psi$ is defined as
$$
\sigma_\psi^2 (\hat A) = \langle (\hat A - \langle \hat A \rangle_\psi I)^2 \rangle_\psi = 
\langle \hat A^2 \rangle_\psi - \langle \hat A \rangle^2_\psi.
$$

Let $\xi$ be a classical random variable. Its dispersion is given 
by $\sigma_P^2 (\xi) = E_P (\xi-E_P \xi)^2 = E_P \xi^2 - (E_P \xi)^2,$
where $P$ is a probability measure. 

\medskip

{\bf Lemma 1.} {\it Let operator $\hat A$ be symmetric and let $\psi$ be a pure state. 
Then
\begin{equation}
\label{c1}
{\rm Tr} (D_{\psi} \hat A)^2 = {\rm Tr} \hat A^2 + 2 \langle \hat A^2 \rangle_\psi + \langle \hat A\rangle_\psi^2
\end{equation}
where $\rho_\psi = \psi \otimes \psi$ and operator $D_{\psi}$ is defined by (\ref{b2}).}

{\bf Proof.} We have (for an orthonormal basis)
$$
\sum_k \langle D_{\psi} \hat A e_k \mid \hat A D_{\psi} e_k\rangle =
\sum_k \langle (I + \rho_\psi) \hat A e_k \mid \hat A (I + \rho_\psi) e_k\rangle
$$
$$
= \sum_k \langle \hat A e_k \mid \hat A e_k\rangle +  \sum_k \langle \hat A e_k \mid \hat A \rho_\psi e_k\rangle + 
\sum_k \langle \rho_\psi \hat A e_k \mid \hat A  e_k\rangle 
$$
$$
+ \sum_k \langle \rho_\psi \hat A e_k \mid \hat A \rho_\psi e_k\rangle  
$$
$$
={\rm Tr} \hat A^2 + 
\sum_k \langle \hat A e_k \mid \langle e_k \mid \psi\rangle  \hat A \psi\rangle + 
\sum_k \langle \langle \hat A e_k \mid \psi\rangle \psi \mid \hat A e_k\rangle
$$
$$
+\sum_k \langle \langle \hat A e_k \mid \psi\rangle \psi \mid \langle e_k\mid \psi\rangle \hat A \psi\rangle = {\rm Tr} \hat A^2 +
$$
$$
\sum_k \langle \psi \mid e_k\rangle \langle e_k \mid \hat A^2 \psi\rangle + \sum_k \langle \hat A \psi \mid e_k\rangle \langle e_k\mid
 \hat A \psi\rangle + 
\sum \langle \psi \mid \hat A \psi\rangle  \langle  e_k\mid \hat A \psi\rangle \langle  \psi\mid e_k \rangle =
$$
$$
{\rm Tr} \hat A^2 + 2 \langle \hat A^2 \psi \mid \psi\rangle + \langle \hat A \psi\mid \psi\rangle^2.
$$

We recall once again 
that each pure quantum state $\psi$ determines the density operator $\rho_\psi.$ The latter determines the covariance 
operator $D_{\psi}$ by (\ref{b2}) of the Gaussian measure $\mu_{D_{\psi}}.$ To simplify notations, we will use the symbol 
$\mu_\psi$ for this measure. Thus   $\mu_\psi$ is the prequantum Gaussian distribution corresponding to the pure 
quantum state $\psi.$ It describes prequantum random field of  ``quantum system coupled to vacuum thermostat.''
We remark that, in particular, our activity is on translation of the operator language of the traditional 
quantum formalism to the language of traditional probability theory. 

We will use the following result on the Gaussian integral of the product of two quadratic forms 
on the complex Hilbert space \cite{CS}:

\medskip

{\bf Lemma 2.} {\it Let $\mu$ be a Gaussian measure with the  
covariance operator $D$ and let $\widehat A_i$ be self-adjoint operators $i= 1,2.$ Then}
\begin{equation}
\label{YY1} E_\mu f_{A_1} f_{A_2}\equiv \int_{H} f_{A_1}( \phi) f_{A_2}(\phi)   d\mu (\phi) = {\rm Tr} D
\widehat A_1 {\rm Tr} D \widehat A_2 + {\rm Tr} D \widehat A_2 D
\widehat A_1.
\end{equation}

\medskip

{\bf Theorem.} {\it Let conditions of Lemma 1 hold. Then}
\begin{equation}
\label{c2}
E_{\mu_\psi} f_A^2 = (E_{\mu_\psi} f_A)^2 + 
{\rm Tr} \hat A^2 + 2 \langle \hat A^2 \rangle_\psi + \langle \hat A \rangle_ \psi^2.
\end{equation}

By (\ref{c2}):
\begin{equation}
\label{c2A}
\sigma_{\mu_\psi}^2 (f_A) = 
{\rm Tr} \hat A^2 + 2 \sigma_\psi^2 (\hat A) + 3\langle A \rangle_\psi^2.
\end{equation}
We remark that the classical dispersion $\sigma_{\mu_\psi}^2 (f_A)$ is always larger than the quantum dispersion  $\sigma_\psi^2 (\hat A)$
 -- since in the last one we ignore the dispersion of vacuum fluctuations. 
Moreover, the classical dispersion is larger than the dispersion produced by vacuum fluctuations
which is given by ${\rm Tr} \hat A^2.$ Really, we have
$$
E_{\mu_I} f_A^2= (\rm{Tr}\; \hat{A})^2 + \rm{Tr}\; \hat{A}^2.
$$ 
By taking into account that $E_{\mu_I} f_A=\rm{Tr}\; \hat{A}$ 
we get $\sigma^2_{\mu_I}(f_A)= \rm{Tr}\; \hat{A}^2.$

Let now $\langle A \rangle_\psi =0.$ Then 
$$
\sigma_\psi^2 (\hat A) = 
\frac{1}{2} [\sigma_{\mu_\psi}^2(f_A) -  {\rm Tr} \hat A^2] = \frac{1}{2} [\sigma_{\mu_\psi}^2(f_A) 
- \sigma^2_{\mu_I}(f_A)] \equiv \Gamma_{\mu_\psi}(f_A).
$$
Thus the QM-dispersion can be obtained as regularization, shift by ${\rm Tr} \hat A^2,$ 
of the classical dispersion (as to the factor 1/2) -- by ignoring the contribution of vacuum fluctuations. 
We remark that {\it equality of the quantum dispersion to zero is nothing else than the reduction of the classical dispersion 
to the dispersion of vacuum fluctuations -- the dispersion of the irreducible background noise.} 

Consider two quantum observables, symmetric operators $\hat A_1$ and $\hat A_2;$ 
in general noncommutative: $[ \hat A_1, \hat A_2] \ne 0.$
Take a pure quantum state $\psi.$ We can always assume that these observables have zero averages in this state: 
\begin{equation}
\label{*}
\langle \hat A_i \rangle_\psi =  \langle \hat A_i \psi \mid \psi \rangle = 0, i=1,2.
\end{equation}
If $\langle \hat A_i \rangle_\psi  \not= 0,$ we just consider shifted observables: 
\begin{equation}
\label{*1}
\hat C_i = \hat A_i - \langle \hat A_i \rangle_\psi I.
\end{equation}
Since the QM dispersions satisfy the Schr\"odinger-Robertson inequality\footnote{
For position and momentum, it was found by
Schr\"odinger and Robertson, see the full story in \cite{DOD}.}:
\begin{equation}
\label{d1}
\sigma_\psi^2 (\hat{A}_1) \sigma_\psi^2 (\hat{A}_2) \geq \frac{1}{4} \vert 
\langle [\hat A_1, \hat A_2]\rangle_\psi \vert^2,
\end{equation}
the dispersions of the corresponding PCSFT variables shifted by the background white noise  satisfy its classical counterpart:
\begin{equation}
\label{d2}
\Gamma_{\mu_\psi}(f_{A_1}) \Gamma_{\mu_\psi}(f_{A_2}) \geq \frac{1}{4} \vert 
\langle [\hat A_1, \hat A_2]\rangle_\psi \vert^2. 
\end{equation}
We now represent even the right-hand side as classical (PCSFT) average. Set 
$$\hat{K}= [\hat A_1, \hat A_2].
$$  
It is a skew-symmetric operator. By applying (\ref{BBPP}) and taking into account the remark that this formula is valid not only 
for quantum observables, but even arbitrary linear operators, we obtain: 
$$
E_{\mu_\psi} f_K = \rm{Tr} \; \rho_\psi  \hat K + \rm{Tr} \; \hat K.
$$
Thus
$$
 \langle \hat K \rangle_\psi=E_{\mu_\psi} f_K - \rm{Tr} \; \hat K.
 $$
Inequality (\ref{d2}) can be written in purely classical terms:
\begin{equation}
\label{d3}
\Gamma_{\mu_\psi}(f_{A_1}) \Gamma_{\mu_\psi}(f_{A_2}) \geq \frac{1}{4} \vert  E_{\mu_\psi} f_K - \rm{Tr} \; \hat K \vert^2. 
\end{equation}
Thus noncommutativity has its trace even in PCSFT.  In contrast to QM, Schr\"odinger-Robertson PCSFT-inequality 
does not have drastic consequences. The main difference is the presence of the shift of the dispersion.
This shift which is by the way produced by the contribution of fluctuations of vacuum makes the restriction on 
the product of the dispersions not so rigid as it is in QM. However, we remind that even 
the conventional Schr\"odinger-Robertson inequality has a statistical interpretation which has nothing to do with incompatibility of 
observables (in the sense of impossibility of joint measurement). This statistical interpretation is due to Margenau and Ballentine,
see \cite{BL1}--\cite{BL3}.  

\medskip

{\bf Conclusion} {\it Prequantum field theory contains a counterpart of the Heisenberg uncertainty relation - 
the Schr\"odinger-Robertson like inequality (\ref{d3}). This classical analog of the 
Schr\"odinger-Robertson inequality provides an estimate from 
below not for the dispersions, but for the shifted (by contribution of vacuum fluctuations) dispersions. Such an estimate is not so rigid 
as the one provided by the Schr\"odinger-Robertson inequality for the quantum dispersions. It may be that paradoxic 
consequences of the Heisenberg's  uncertainty principle were induced by neglecting of the mentioned shift.}

\medskip

I would like to thank L. Ballentine who explained me the statistical interpretation of Heisenberg's uncertainty relation.

\end{document}